\documentclass[aps,prl,twocolumn,preprintnumbers,superscriptaddress,showpacs,floatfix]{revtex4-1}

\usepackage{hyperref}
\usepackage{color}
\usepackage{graphicx}
\usepackage{amsmath}	
\usepackage{siunitx}
\graphicspath{
  {./},
}

\usepackage{amstext}
\usepackage{array}

\usepackage{xcolor}
\usepackage{xspace}	

\begin{document}

\title{Searching for the QCD critical point via the rapidity dependence of cumulants}

\author{Jasmine Brewer}
\affiliation{Center for Theoretical Physics, Massachusetts Institute of Technology, Cambridge, MA 02139, USA}
\author{Swagato Mukherjee}
\affiliation{Physics Department, Brookhaven National Laboratory, Upton, NY 11973, USA}
\author{Krishna Rajagopal}
\affiliation{Center for Theoretical Physics, Massachusetts Institute of Technology, Cambridge, MA 02139, USA}
\author{Yi Yin}
\affiliation{Center for Theoretical Physics, Massachusetts Institute of Technology, Cambridge, MA 02139, USA}

\date{\today}

\begin{abstract}

The search for a possible critical point in the QCD phase diagram is ongoing in heavy ion collision experiments at RHIC which scan the phase diagram by scanning the beam energy; a coming upgrade will increase the luminosity and extend the rapidity acceptance of the STAR detector. 
In  fireballs produced in RHIC collisions, the baryon density depends on rapidity. By employing Ising universality together with a phenomenologically motivated freezeout prescription, we show that the resulting rapidity dependence of cumulant observables sensitive to critical fluctuations is distinctive. The dependence of the kurtosis (of the event-by-event distribution of the number of protons) on rapidity near mid-rapidity will change qualitatively if a critical point is passed in the scan. Hence, measuring the rapidity dependence of cumulant observables can enhance the prospect of discovering a critical point, in particular if it lies between two energies in the beam energy scan.

\end{abstract}

\preprint{MIT-CTP/4998}
\maketitle

A central goal of heavy ion collision experiments is to map the QCD phase diagram as a function of temperature $T$ and baryon chemical potential $\mu_B$~\cite{Akiba:2015jwa,Geesaman:2015fha,Busza:2018rrf}. 
At zero $\mu_B$, the phase diagram features a continuous crossover from quark-gluon plasma (QGP) to ordinary hadronic matter as a function of decreasing $T$~\cite{Karsch:2001vs,Aoki:2006we,Borsanyi:2013bia,Bazavov:2014pvz,Bhattacharya:2014ara}.
Increasing $\mu_B$ corresponds to doping the QGP with an excess of quarks over antiquarks, and it is an open question whether the crossover becomes a sharp first order phase transition beyond some critical point~\cite{Luo:2017faz,Busza:2018rrf}. 
At nonzero $\mu_B$ where lattice calculations become extremely difficult~\cite{deForcrand:2010ys,Ding:2015ona},
there are no first-principles theoretical calculations which provide reliable guidance as to the whether there is a critical point in the phase diagram of QCD, or its location if it does exist~\cite{Rajagopal:1999cp,Rajagopal:2000wf,Stephanov:2004wx,Stephanov:2007fk}. 
Model calculations 
suggest the existence of a critical point, but disagree wildly on its location in the $(\mu_B,T)$ plane~\cite{Stephanov:2004wx,Stephanov:2007fk}. 
Reducing the beam energy increases the $\mu_B$ of the QGP produced in a heavy ion 
collision~\cite{Bearden:2003hx,Cleymans:2005xv,Andronic:2017pug,Busza:2018rrf} 
(principally because lower energy collisions make less entropy but also because they deposit more of their baryon number in the plasma) but it also reduces the temperatures achieved. So, these experiments can scan the crossover (and potentially critical) regime of the phase diagram out to some value of $\mu_B$ corresponding to the lowest energy collisions that reach the crossover (critical) temperature~\cite{Akiba:2015jwa,Geesaman:2015fha}.
If a critical point is located in the regime that is within reach, it may be detected experimentally. The search for a critical point in the phase diagram of QCD at the Relativistic Heavy Ion Collider (RHIC) is currently underway, with collisions at energies ranging from $\sqrt{s}=200$~AGeV down to $\sqrt{s}=7.7$~AGeV, producing fireballs that freeze out with chemical potentials in the range  $25~{\rm MeV} \lesssim \mu_B \lesssim 400$~ MeV~\cite{Cleymans:2005xv,Andronic:2017pug}.
Phase I of the RHIC beam energy scan (BES-I) was completed in 2014, with no signs of a critical point for $\mu_B<200$~MeV and  
with tantalizing but inconclusive results at larger $\mu_B$, in collisions with 
$19.6~{\rm AGeV}\geq \sqrt{s} \geq 7.7$~AGeV~\cite{Adamczyk:2013dal,Luo:2015ewa,Luo:2015doi,Akiba:2015jwa,Geesaman:2015fha,Luo:2017faz}.  Phase 2 of the scan (BES-II), to begin in 2019~\cite{Luo:2015doi,Luo:2017faz}, will focus on this regime with increased luminosity and consequently much higher statistics. 
One of the improvements planned before BES-II is an upgrade of the inner Time Projection Chamber (iTPC) at STAR, which will extend its 
rapidity acceptance for protons from $|y|<0.5$ in BES-I to $|y|<0.8$ in BES-II~\cite{Wang:2014mda}.

The energy of a heavy ion collision sets the initial $T$ and $\mu_B$ of the QGP which is created, with lower energy collisions being more baryon-rich. The QGP then follows a trajectory in the $(T,\mu_B)$ plane as it expands and cools.
If there is a critical point in the QCD phase diagram within the range of $\mu_B$ which is accessible in the Beam Energy Scan (BES), then at 
some collision energies the fireball produced may pass through or near the critical region, while
at
higher (lower) collision energies the fireball produced will pass the critical point 
on the low (high) $\mu_B$ side.
The theoretical challenge is to describe the unique signatures of this scenario which would be observable in data from the BES.

A critical point in a thermodynamic system is characterized by an enhanced correlation length. Although the correlation length itself is not observable, because the critical order parameter $\sigma$ couples 
to all hadrons 
the $n$'th cumulant moments $\kappa_n$ 
of the event-by-event distribution of the measured multiplicity $N$ of various particle species, for example $\kappa_4[N] = \langle (\delta N)^4 \rangle - \langle (\delta N)^2 \rangle^2$, scale with powers of the correlation 
length $\xi$ near a critical point~\cite{Stephanov:2008qz}. Protons couple more strongly 
to $\sigma$ than pions or kaons, making cumulants of the proton multiplicity good observables with which to look for critical fluctuations~\cite{Hatta:2003wn,Athanasiou:2010kw}. 
The non-Gaussian 
cumulants $\kappa_3[N] =\langle (\delta N)^3 \rangle \sim \xi^{9/2}$ and $\kappa_4[N] \sim \xi^7$ 
scale with higher powers of the correlation length than the Gaussian cumulants and are therefore more sensitive to critical behavior~\cite{Stephanov:2008qz,Athanasiou:2010kw}. 
Furthermore, an analysis that is valid for any critical point in the same (3d Ising) universality class has shown that $\kappa_4$ will also change sign near a QCD critical point \cite{Stephanov:2011pb,Stephanov:2011zz}. Non-monotonic behavior and a sign change of the fourth cumulant as a function of the beam energy are characteristic signatures of the presence of a critical point which can be searched for in the RHIC BES. 

The dependence of fluctuation measures on the total rapidity acceptance has been studied before~\cite{Ling:2015yau,Bzdak:2016sxg} upon assuming that $\mu_B$, and hence $\xi$ and $\kappa_4$,  do not depend on rapidity. However, the baryon density does depend non-trivially on rapidity at RHIC BES energies (see e.g.~Refs.~\cite{Becattini:2007ci,Shen:2017bsr}). Since the correlation length and fluctuations become very large near a critical point, the rapidity dependence of the baryon density gives rise to a strong, non-trivial rapidity dependence of the cumulants near a critical point which was not incorporated in previous work. 
Furthermore, since each rapidity is associated with a different value of $\mu_B$ and therefore probes a different part of the critical region on the phase diagram, we shall see that integrating over the full rapidity acceptance averages out interesting features in the cumulants which are characteristic of critical behavior. Instead, we propose that binning the cumulants in rapidity gives a more crisp picture of the critical regime, and demonstrate that the rapidity dependence of these binned cumulants near mid-rapidity will change qualitatively if a critical point is passed in the BES. 
We therefore propose this observable as a complementary means by which to observe the presence of a critical point at the BES. In particular, it provides
a new, and distinctive, signature by which to determine whether downward steps in the collision energy take us past a critical point in the phase diagram.

{\bf Rapidity-dependence of $\mu_B$:} In this Letter, we  illustrate the effect that the rapidity-dependence of $\mu_B$ at RHIC BES energies has on the rapidity-dependence of cumulants, and propose using this  toward discovering (or ruling out) a critical point using RHIC BES data.

Since the baryon density at freezeout is symmetric in the spacetime rapidity $y_s$ for symmetric heavy-ion collisions, 
for small $|y_s|$ the deviation  from boost invariance takes the form
\begin{equation}
	\label{eq:muBofeta}
\mu_B(y_s) \sim \mu_{B,0} + \alpha\, y_s^2\,,
\end{equation}
with $\mu_{B,0}$ and $\alpha$ constants that depend on the beam energy $\sqrt{s}$. We shall use this form for illustrative purposes, noting of course that it cannot be relied upon at large $|y_s|$.
As we have discussed, the basis of the BES is that downward steps in $\sqrt{s}$ yield upward steps in $\mu_{B,0}$.  For illustrative purposes, we shall pick three values of $\mu_{B,0}$ within the BES range, and see what happens if these steps were to happen to take us past a possible critical point.
The value of $\alpha$ has been measured in SPS collisions with $\sqrt{s}=17.3$~AGeV, where $\alpha=50$~MeV~\cite{Becattini:2007ci}.   At this (and all higher, and some lower) beam energies, $\alpha>0$ because the baryon number density is peaked at roughly two units of rapidity below the beam rapidity, meaning that it is less at $y_s=0$ than at larger $|y_s|$~\cite{Bearden:2003hx,Busza:2018rrf}.  In AGS collisions with $\sqrt{s}=5.5$~AGeV, though, the beam rapidity is low enough that the baryon number density is peaked at $y_s=0$, and $\alpha<0$~\cite{Bearden:2003hx}.   The $\sqrt{s}$ at which $\alpha$ changes from positive to negative is not known, but is likely near the lower end of the BES range.
Ultimately, measurements of the ratios of the mean particle number distribution for different species 
from the RHIC BES should be used to measure how $\mu_B$ at freezeout depends on $y_s$ at each BES collision energy, and hence to determine the value of $\alpha$ at each energy. For illustrative purposes here, we shall investigate the consequences of choosing
$\alpha=50$~MeV at each of our three values of $\mu_{B,0}$ as well as checking how things change if we choose $\alpha=-50$~MeV instead at our largest value of $\mu_{B,0}$.

{\bf Cumulants in the critical regime:} Order parameter fluctuations near a critical point induce fluctuations in the event-by-event particle multiplicities. Throughout this work, we will consider the cumulants of protons, as these are expected to be most sensitive to critical fluctuations~\cite{Athanasiou:2010kw}. From previous work~\cite{Stephanov:2008qz,Athanasiou:2010kw,Stephanov:2011pb,Ling:2015yau}, the contribution to the fourth cumulant of the proton multiplicity distribution 
coming from critical fluctuations (denoted by the subscript $\sigma$) takes the form
\begin{equation}
	\label{eq:kappa4}
	\kappa_4[N]_\sigma 
	= \int_{\bf x} K_4 \,
	\xi^7 T^{2} \left( g \int_{\bf p} \frac{ \chi_{\bf p}}{\gamma_{\bf p}} \right)^4, 
\end{equation}
where the ${\bf x}$-integral is a spacetime integral over the freezeout hypersurface, 
where $T$, $\mu_B$ and consequently $\xi(\mu_B,T)$  and $K_4(\mu_B,T)$ (proportional to the kurtosis of the event-by-event distribution of the fluctuating order parameter, see below) can take on different values at different points on the freezeout hypersurface,
where $g$ is the $\sigma$-proton-proton coupling which we set to the same benchmark value $g=7$ used in Ref.~\cite{Athanasiou:2010kw}, 
where the ${\bf p}$-integral is a momentum-space integral over the protons at the point ${\bf x}$,
where $\chi_{\bf p}= f_{\bf p}(1-f_{\bf p})/T$ if we assume local equilibrium, with $f_{\bf p}$ the Fermi-Dirac distribution boosted by the radial flow velocity at the point ${\bf x}$,
and where $\gamma_{\bf p}=\sqrt{{\bf p}^2+m^2}/m$, with $m$ the proton mass.
We follow Ref.~\cite{Ling:2015yau} and use a blast wave model to obtain the radial flow velocity and freezeout hypersurface, taking the freezeout curve in the $(T,\mu_B)$ plane from the fit to experimental data found in Ref. \cite{Cleymans:2005xv,Andronic:2017pug}. 
Following Ref.~\cite{Ling:2015yau}, we shall make the approximation $\chi_{\bf p}\approx f_{\bf p}/T$ and use the Boltzmann distribution for $f_{\bf p}$, allowing us to do some of the integrals analytically.

\begin{figure*}[t]
\centering
\includegraphics[width=18cm]{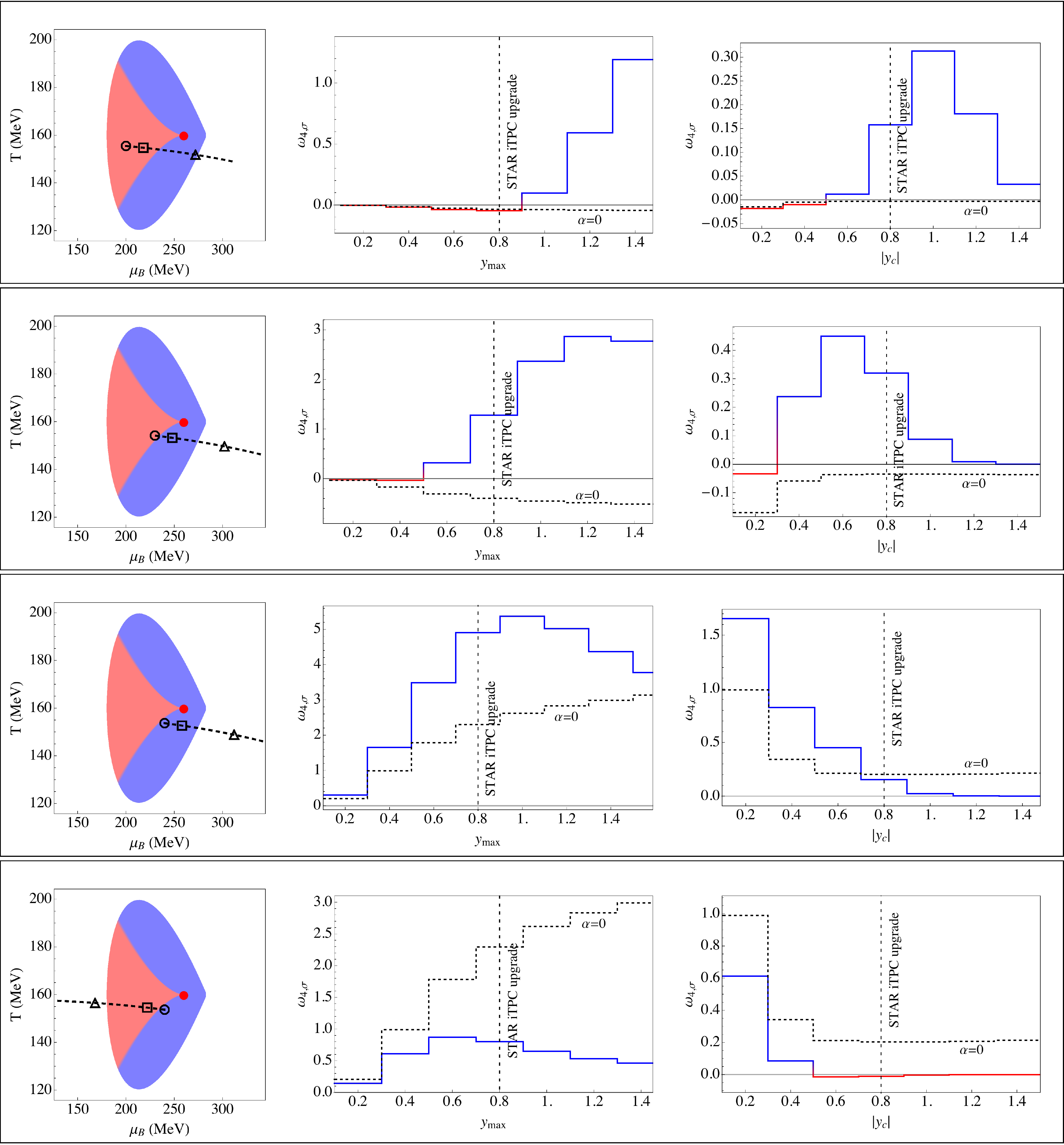}
\par
\caption{\label{fig:4-panel}
In the left column we see that we have assumed the existence of a critical point (red dot) at $(\mu_B^c,T)=(260,160)$~MeV whose critical region, bounded by the contour where $\xi =1$~fm, 
is colored red and blue. The colors denote the sign of $\omega_{4,\sigma}$, with $\omega_{4,\sigma}>0$ in blue and $\omega_{4,\sigma}<0$ in red. 
Different rows correspond to different assumptions for where on the phase diagram a heavy ion collision freezes out, {\it cf.}~collisions with varying beam energy. The black circles show where freezeout occurs at mid-rapidity, 
from top to bottom with $\mu_{B,0} = 200$, 230, 240, 240~MeV. The black dashed curves show how the freezeout conditions change with increasing spacetime rapidity, with the circle, square, and triangle  indicating freezeout at $y_s = 0$, 0.6, and 1.2, respectively. 
In the top three rows, we have chosen $\alpha = \SI{50}{MeV}$ (see eq.~\ref{eq:muBofeta}) while for the bottom row we have chosen $\alpha = \SI{-50}{MeV}$. 
The middle column shows how $\omega_{4,\sigma}$ computed for a rapidity acceptance $|y|<y_{\rm max}$
depends on $y_{\rm max}$.
The right column shows how $\omega_{4,\sigma}$ computed in a pair of bins with width $\Delta y=0.4$ centered at $\pm y_c$ depends on $y_c$.  The results in the middle column sum over a wide range of rapidities (with $|y|$ between 0 and $y_{\rm max}$) which freezeout with a range of $\mu_B$,  meaning that features from the left column are more directly visible in the right column than in the middle.
In both the center and right columns, the black dotted lines show $\omega_{4,\sigma}$ with $\alpha=0$, {\it i.e.}~what would have been obtained if $\mu_B=\mu_{B,0}$, denoted by the black circles in the left column, everywhere.
The results shown in the right and middle column should not be taken as quantitative predictions since they depend on the many assumptions that we made for illustrative purposes; they are illustrative of qualitative features to be expected in the rapidity-dependence of cumulants if steps in beam energy take us past a critical point. 
}
\end{figure*}

The shape of the dependence of $K_4$ and $\xi$ on $\mu_B$ and $T$ are governed by
universal properties of critical fluctuations.
A critical point in the QCD phase diagram, if it exists, is known to be in the same universality class as the 3d Ising model \cite{Rajagopal:1992qz,Berges:1998rc,Halasz:1998qr,Stephanov:1998dy,Stephanov:1999zu}. The mapping of the Ising variables $(r,h)$ onto the QCD variables $(\mu_B,T)$ is not universal, but for illustrative purposes
we employ the widely-used assumption \cite{Berdnikov:1999ph} that the Ising $r$ axis (and hence the line of first order transitions)
is parallel
to the QCD $\mu_B$ axis, and the Ising $h$ axis is parallel to the QCD $T$ axis.  For illustrative purposes, we shall place a hypothetical QCD critical point at $\mu_B=260$~MeV, $T=160$~MeV.
The 3d Ising universality then determines $K_4$ at some point away from the critical point in terms of the direction in which that point lies in the $(\mu_B,T)$ plane, and $\xi$ in terms of this angle, the distance away from the critical point, and one non-universal parameter whose choice determines the contour on the phase diagram where $\xi=1$~fm, as illustrated in Fig.~\ref{fig:4-panel}. Because once $\xi$ is less than 1~fm the magnitude of $\kappa_4 \propto \xi^7$ is negligible, for simplicity we set $\xi=0$ outside the critical regime.  (For details, see Refs.~\cite{ZinnJustin,Athanasiou:2010kw,Stephanov:2011pb,Parotto:2018pwk}.)

Following Ref.~\cite{Ling:2015yau},
we cast the momentum integration
in terms of the momentum-space rapidity $y$ and transverse momentum $p_\perp$, which are measured in experiment: 
\begin{equation}
	\label{eq:kappa4_mom}
\int_{\bf p} \frac{1}{\gamma_{\bf p}}\rightarrow \frac{2m}{(2\pi)^3}\int_{y_c -\Delta y/2}^{y_c + \Delta y/2} dy \int_{p_{\mathrm{min}}}^{p_{\mathrm{max}}} p_\perp dp_\perp \int_0^{2\pi} d\psi\, .
\end{equation}
We have introduced a finite acceptance in both rapidity and transverse momentum.
We will keep  $p_{\rm min} = \SI{0.4}{GeV}$ and $p_{\rm max}=\SI{2}{GeV}$ throughout. We shall compute $\kappa_4$ using two different kinds of rapidity cuts, either varying  $\Delta y$ with $y_c=0$, in which case $|y|<y_{\rm max}\equiv \Delta y/2$, or varying $y_c$ with fixed bin width $\Delta y$.

\begin{figure}
\centering
\includegraphics[width=8cm]{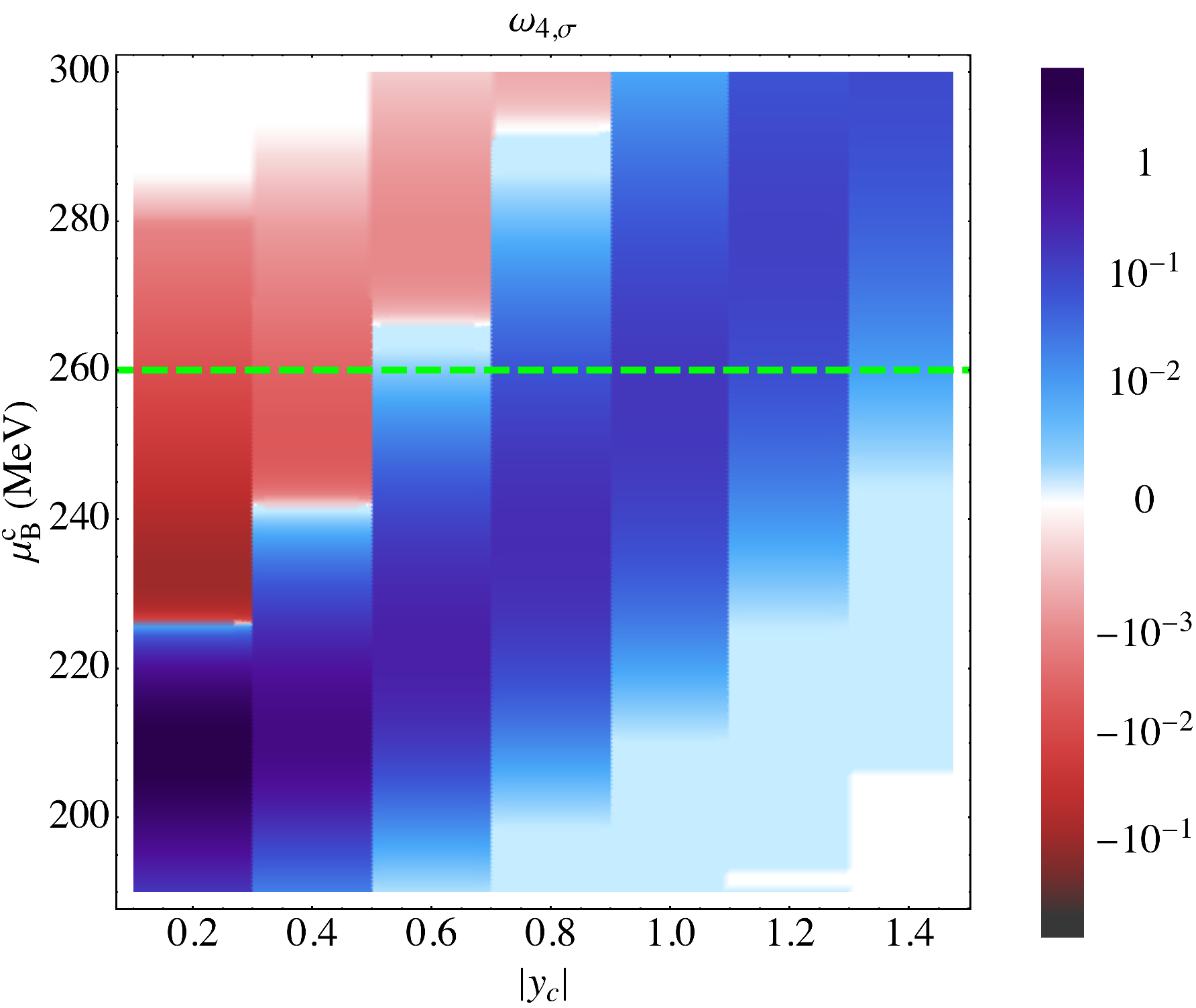}
\par
\caption{We illustrate the behavior of $\omega_{4,\sigma}$ when the freezeout conditions are as in the
first row of Figure \ref{fig:4-panel} with $\mu_{B,0}=\SI{200}{MeV}$ and $\alpha=50$~MeV, but where here we let the location of the critical point 
range from $\mu_B^c=190$~MeV to $\mu_B^c=300$~MeV.   For each value of $\mu_B^c$, namely for each horizontal slice across the figure, color indicates the value of $\omega_{4,\sigma}$ as a function 
of $|y_c|$, with $\Delta y=0.4$ fixed as in Fig.~\ref{fig:4-panel}. The slice indicated by the green dashed line corresponds 
to the top-right panel of Fig.~\ref{fig:4-panel}. 
For $\mu_B^c < \mu_{B,0}$
($\mu_B^c > \mu_{B,0}$), $\omega_{4,\sigma}$ decreases (increases) with increasing $|y_c|$.
}
\label{fig:density-plot}
\end{figure}

To simplify the interpretation of our results, we shall show the critical contribution to the cumulants normalized by the average number of protons, 
$
	\omega_{4,\sigma} \equiv \kappa_{4}[N]_{\sigma} / \langle N \rangle	
$.
This cumulant ratio 
has the advantage that if the background (noncritical) contribution were Poisson-distributed it would contribute $\omega_{4,\sigma} =1$, 
meaning that our results in Figures \ref{fig:4-panel} and \ref{fig:density-plot} should be interpreted as critical contributions to be added to a background of order 1.

{\bf Results and conclusions}: In this Section, we demonstrate that the rapidity-dependence of $\mu_B$ makes the rapidity dependence of cumulants sensitive to critical fluctuations in a way that yields distinctive, qualitative, observable consequences. In Fig.~\ref{fig:4-panel} we first compute the dependence of the cumulant ratio $\omega_{4,\sigma}$ on the total rapidity acceptance $y_{\rm max}$.  This dependence was studied previously in Ref.~\cite{Ling:2015yau} upon assuming that $\mu_B$ itself is constant in rapidity; we find striking consequences of the rapidity-dependence of $\mu_B$.  Next, motivated by the expanded rapidity coverage that the STAR iTPC upgrade will bring, we compute  $\omega_{4,\sigma}$ for bins in rapidity, something that has not been considered previously.   We find that the rapidity dependence of the cumulant ratio is a sensitive and
interesting probe of critical behavior.

In Fig.~\ref{fig:4-panel} we consider a hypothetical set of scenarios 
motivated by the possibility that there may be a QCD critical point within the energy
range to be explored by the RHIC BES. We imagine a critical point at $\mu_B^c=260$~MeV, and in the first three rows of the figure we consider heavy ion collisions with three decreasing values of the beam energy such that freezeout at mid-rapidity occurs at $\mu_{B,0}=200$, 230, 240~MeV.  In all three rows, we choose 
$\alpha=50$~MeV, corresponding to the measured value from SPS collisions with $\sqrt{s}=17.3$~GeV and $\mu_{B,0}=237$~MeV.  Because a real critical point may lie at larger $\mu_B^c$ than this, where $\alpha$ may become negative, in the fourth row we flip the sign of $\alpha$. The right column of Figure \ref{fig:4-panel} shows $\omega_{4,\sigma}$ binned in rapidity bins of width $\Delta y = 0.4$ centered around $y = \pm y_c$, an observable which to our knowledge has not been considered before. This is a more sensitive observable to the unique features of 
critical behavior than the dependence on the total rapidity acceptance in the center column because it isolates contributions coming from more similar values of $\mu_B$, and the correlation length and other features of the critical regime are sensitive to $\mu_B$ near $\mu_B^c$. 
We see many interesting qualitative features in the rapidity dependence of $\omega_{4,\sigma}$. For example, if $\mu_{B,0}$ is in the red region, where $\omega_{4,\sigma}$ is negative and relatively small in magnitude, larger and positive contributions to $\omega_{4,\sigma}$ can be found at larger rapidity.
This can be seen in the middle panels of the first and second rows, but it is much more striking in the right panels, indicating the value of binning in rapidity. On the other hand,  if $\mu_{B,0}$ lies in the blue region, in the right column the largest value of $\omega_{4,\sigma}$ is obtained for the bin centered at $y=0$, with $\omega_{4,\sigma}$ decreasing with increasing rapidity while staying positive if $\alpha>0$ as in the third row or decreasing with increasing rapidity while becoming negative if $\alpha<0$ as in the fourth row. Both the sign change and the non-monotonic behavior in  $\omega_{4,\sigma}$, as a function of the rapidity acceptance in the center panel of Fig.~\ref{fig:4-panel} and even more so as a function of the rapidity bin in the right panel of Fig.~\ref{fig:4-panel}  
are new results of this work.  They arise from the rapidity dependence of $\mu_B$ at freezeout in collisions at RHIC BES energies, and provide distinctive signatures if decreasing the beam energy in this scan takes $\mu_{B,0}$ past a critical point.

To complement Fig.~\ref{fig:4-panel}, Fig.~\ref{fig:density-plot} shows the cumulant ratio $\omega_{4,\sigma}$ binned in rapidity for a fixed beam energy (fixed $\mu_{B,0}$) as the location of the critical point $\mu_B^c$ is changed. There are several features of binning the cumulants in rapidity which we believe will make doing so an important way to probe the critical region, if a critical point is discovered in the RHIC BES.
First, $\omega_{4,\sigma}$ increases with $|y_c|$ if freezeout at mid-rapidity occurs at a $\mu_{B,0}$ that is well below $\mu_{B}^c$, in the red region, whereas  it will decrease with $|y_c|$ if $\mu_{B,0}$ is closer to or larger than $\mu_B^c$, in the blue region.  This remains true even if $\alpha$ changes sign, as demonstrated in the bottom row of Fig.~\ref{fig:4-panel}.
Furthermore, a sign change in $\omega_{4,\sigma}$ as a function of $y$ will be easier to see upon binning in $|y_c|$ since not doing so, as in the middle panels, can obscure it by mixing data from different regions in rapidity.   Even in cases where the sign change in $\omega_{4,\sigma}$ is visible in the middle column, as in the first and second rows, it happens at a lower rapidity in the right panel than in the middle panel, making it more feasible to observe at STAR via binning in $|y_c|$.

We conclude that the rapidity-dependence of $\mu_B$ at RHIC energies may result in qualitative signatures of critical fluctuations manifest in the rapidity-dependence of the cumulant ratio $\omega_{4,\sigma}$. 
Complementary to scanning  the phase diagram by taking steps in beam energy, the rapidity dependence of $\mu_B$ provides additional scans of small regions of the phase diagram. We have seen that non-monotonicity and a sign change of the critical contribution to $\omega_{4,\sigma}$ as a function of rapidity will arise if the BES includes energies on both sides of a critical point. Binning the cumulants in rapidity provides a sensitive probe of these effects. Signatures of critical behavior
in the $\sqrt{s}$-dependence of $\omega_{4,\sigma}$
can therefore be cross-checked by looking for qualitative changes in the rapidity-dependence of $\omega_{4,\sigma}$ 
between nearby beam energies. 
We have made arbitrary choices at many points, for illustrative purposes. 
A future quantitative study should include investigation of changes to these choices, as well as an
investigation of consequences of variations in the baryon density at a given rapidity, for example along the lines of Ref.~\cite{Shen:2017bsr}. We note however that if the value of $\alpha$ is determined from experimental data as in Ref.~\cite{Becattini:2007ci} this will incorporate the most important such consequence. Future studies should also include an analysis of the 
quantitative effects of non-equilibrium dynamics on
the values of $\sqrt{s}$ or $y_c$ at which the qualitative features we have found 
occur~\cite{Berdnikov:1999ph,Mukherjee:2015swa,Mukherjee:2016kyu,Sakaida:2017rtj}.

\begin{acknowledgments}
\paragraph{\bf Acknowledgements:}

We are grateful to Ulrich Heinz, Jiangyong Jia, Roy Lacey, Xiaofeng Luo, Jean-Yves Ollitrault, Hannah Petersen, Lijuan Ruan, Bj\"{o}rn Schenke, Chun Shen, Misha~Stephanov and Raju Venugopalan for helpful conversations and Francesco Becattini for a helpful private communication about Ref.~\cite{Becattini:2007ci}.
We are grateful to Wilke van der Schee for help realizing the visualization that we have employed in Fig.~2.
This work was supported by the Office of Nuclear Physics of the U.S. Department of Energy within the framework of the Beam Energy Scan Theory (BEST) Topical Collaboration and under Contracts  DE-SC0011090 (JB, KR, YY) and DE-SC0012704 (SM).

\end{acknowledgments}

\bibliographystyle{apsrev} \bibliography{rapidity-CEP}

\end{document}